\documentclass[twocolumn,10pt,superscriptaddress,aps,pra,nofootinbib]{revtex4-2}
\usepackage[T1]{fontenc}
\usepackage[dvipsnames]{xcolor}
\usepackage{float}
\usepackage{inputenc}
\usepackage{amsthm}
\usepackage{thmtools}
\usepackage{mathtools}
\usepackage{physics}
\usepackage{afterpage}
\usepackage{bm} 
\usepackage{bbm} 
\usepackage{braket}
\usepackage{dsfont}
\usepackage{amsmath}
\usepackage{amssymb}
\usepackage{tikz}
\usetikzlibrary{arrows, positioning}
\usepackage{tikz-cd}
\usepackage{ulem}
\usepackage[linesnumbered,ruled]{algorithm2e}
\SetArgSty{textnormal}

\usepackage{etoolbox}
\apptocmd{\sloppy}{\hbadness 10000\relax}{}{}

\usepackage{silence}
        \DeclareMathOperator{\Mat}{Mat} 
        
        \newcommand*{\defcolon}{\,:\,} 
        \newcommand\scriptin{\raisebox{0.15ex}{$\scriptscriptstyle\in$}} 
        
        \newcommand*{\quater}{\text{Q}} 
        \newcommand*{\rstring}{\mathbf{r}} 
        \newcommand*{\sstring}{\mathbf{s}} 
        \newcommand*{\tstring}{\mathbf{t}} 
        \newcommand*{\cmw}{\Omega} 
        \newcommand*{\bigo}{\mathcal{O}} 

\usepackage[colorlinks]{hyperref}
\hypersetup{
    colorlinks  = true,
    citecolor   = PineGreen,
    linkcolor   = MidnightBlue,
    urlcolor    = PineGreen
}

\begin{document}

\title{Tensorized Pauli decomposition algorithm}

\begin{abstract}
    This paper introduces a novel general-purpose algorithm for Pauli decomposition that employs matrix slicing and addition rather than expensive matrix multiplication, significantly accelerating the decomposition of multi-qubit matrices.
    In a detailed complexity analysis, we show that the algorithm admits the best known worst-case scaling and more favorable runtimes for many practical examples.
    Numerical experiments are provided to validate the asymptotic speed-up already for small instance sizes, underscoring the algorithm's potential significance in the realm of quantum computing and quantum chemistry simulations.
\end{abstract}

\author{Lukas Hantzko}
\affiliation{Institut f\"ur Theoretische Physik, Leibniz Universit\"at Hannover, Germany}
\thanks{Submitted to: \textit{Phys. Scr.}}
\author{Lennart Binkowski}
\email{lennart.binkowski@itp.uni-hannover.de}
\affiliation{Institut f\"ur Theoretische Physik, Leibniz Universit\"at Hannover, Germany}
\author{Sabhyata Gupta}
\affiliation{Institut f\"ur Theoretische Physik, Leibniz Universit\"at Hannover, Germany}

\maketitle

\section{Introduction}\label{section:Introduction}

Pauli matrices are ubiquitous in the realm of quantum physics and hold a fundamental importance in many other fields like quantum computing, information, and simulation.
Along with the $2 \times 2$ identity matrix Pauli matrices form a complete basis, spanning the space of all $2 \times 2$ matrices.
The Pauli group generated by $\sigma^{\{0, 1, 2, 3\}} \coloneqq \{I, X, Y, Z\}$ constitutes the elements of the Clifford group, which define the fundamental gate operations in the circuit model of quantum computers \cite{Nielsen2012QuantumComputationAndQuantumInformation} and are also of utmost importance in the context of measurement-based quantum computing \cite{Raussendorf2003MeasurementBasedQuantumComputationOnClusterStates}.
They also play a crucial role in quantum error correction due to their pivotal importance in the theory of stabilizer codes.
Pauli matrices and their tensorized products -- the Pauli strings -- are used for describing errors in quantum computers -- a recent work especially focuses on Markovian decoherence processes of qubits \cite{DeLeon2022PauliComponentErasingQuantumChannels} -- and for generating stabilizer codes to detect and correct errors \cite{Gottesman1998TheoryOfFaultTolerantQuantumComputation}.
They are essential in quantum simulation too as they are used in the description of Hamiltonians of many physical systems that can be mapped onto spin models in quantum many body physics \cite{Georgescu2014QuantumSimulation}, as well as describing electronic or molecular Hamiltonians in quantum chemistry \cite{McArdle2020QuantumComputationalChemistry}.
For ground state preparation of materials \cite{Babbush2018LowDepthQuantumSimulationOfMaterials}, chemistry problems \cite{Lee2021EvenMoreEfficientQuantumComputationsOfChemistryThroughTensorHypercontraction}, generic Hamiltonians \cite{Sun2024HighPrecisionAndLowDepthEigenstatePropertyEstimationTheoryAndResourceEstimation}, and time-dependent Hamiltonians \cite{Kunold2024VectorizationOfTheDensityMatrixAndQuantumSimulationOfTheVonNeumannEquationOfTimeDependentHamiltonians},
the respective Hamiltonian's Pauli decomposition is crucial for all simulation-based classical and quantum algorithms;
the number of Pauli terms directly dictates the complexity of the simulation task.
In principle, any Hamiltonian -- as it can be expressed as a linear combination of tensor products of Pauli matrices -- can be simulated using a quantum simulator \cite{Feynman1982SimulationPhysicsWithComputers}, leading also to many quantum annealing protocols \cite{Copenhaver2021UsingQuantumAnnealersToCalculateGroundStatePropertiesOfMolecules} that rely on the prior decomposition of the evolutionary Hamiltonian into Pauli strings.
Hamiltonian simulation, and thus Pauli decomposition, further plays a pivotal role in leading quantum algorithms for addressing linear algebra problems such as the Harrow-Hassidim-Lloyd algorithm \cite{Harrow2009QuantumAlgorithmForLinearSystemOfEquations} or quantum singular value transformation \cite{Gilyen2019QuantumSingularValueTransformationAndBeyondExponentialImprovementsForQuantumMatrixArithmetics}.

This operation of Pauli decomposition, however, often comes at a significant computational cost, particularly when dealing with multi-qubit operators.
In the quest to accelerate this task, we introduce the Tensorized Pauli Decomposition (\texttt{TPD}) algorithm.
The \texttt{TPD} algorithm sidesteps the resource-intensive matrix multiplication typically used for Pauli decomposition.
Instead, it leverages the tensor structure of Pauli strings to determine the Pauli weights via recursive matrix slicing to reveal the decomposition tensor factor-wisely.
Furthermore, input from real-world application often comes with rich tensor product structure.
Due to its natural tensorial structure, the \texttt{TPD} algorithm detects these structures automatically, avoiding the potential overhead of verifying certain properties of the Hamiltonian in advance.
The \texttt{TPD} makes use of the following key insight:
Given an $n$-qubit matrix, all the weights corresponding to Pauli strings with the same Pauli matrix in the first tensor factor can be stored in an $(n - 1)$-qubit matrix, computable by adding up two blocks of the original matrix.
Applying the same step to the resultant matrices leads to $(n - 2)$-qubit matrices and so on.
After $n$ steps, the matrices become scalars and reveal the weights of the individual Pauli strings in the original $n$-qubit matrix.

The rest of the article is organized as follows:
In \autoref{section:Preliminaries}, we first introduce the formulation of Pauli decomposition and review the existing algorithms that tackle this task together with their worst-case complexities.
In \autoref{section:Methods}, we give an in-depth formulation of three variants of the \texttt{TPD} algorithm: a recursive, an iterative, and a partial decomposition version.
In \autoref{section:Results}, we deduce the worst-case and several special-case runtimes depending of number of qubits for the \texttt{TPD}.
Most importantly, we establish the best known worst-case complexity for the \texttt{TPD} as well as a strong separation from other algorithms for highly structured input such as the Transverse Field Ising Model Hamiltonian.
Moreover, we present our results from benchmark tests we perform to gauge practical performance of the \texttt{TPD} algorithm compared to existing algorithms, showcasing already a practical speed-up rather than just an asymptotic improvement.
Finally in \autoref{section:ConclusionAndOutlook}, we summarize our theoretical and numerical findings that, together, draw a very promising picture of the \texttt{TPD}'s capabilities.
We also discuss potential next steps to broaden the scope of our method even further.

\begin{figure*}[!th]
    \begin{minipage}[t]{0.47\linewidth}
        \begin{algorithm}[H]
            \caption{Recursive \texttt{TPD}($A$, $L$, $\sstring$=``\,'')}
            \eIf{$\dim(A) = 1 \times 1$}
              {
                append $(\sstring, A =\omega_{\sstring})$ to $L$;
              }
              {
                \For{$\text{t} \in \quater$}{
                    compute $\cmw_{\text{t} \bm{\ast}}$ according to \eqref{equation:CumulativeWeightsDefinition}\;
                    \If{$\cmw_{\text{t} \bm{\ast}} \neq 0$} {
                        call Recursive \texttt{TPD}($\cmw_{\text{t} \bm{\ast}}$, $L$, $\text{t} \sstring$)\;
                    } 
                }
              }
        \end{algorithm}
    \end{minipage}\hfill
    \begin{minipage}[t]{0.47\linewidth}
        \begin{algorithm}[H]
        \caption{Iterative \texttt{TPD}($A$)}
            $S^{1} \coloneqq \{\bm{\ast}\}$\;
            \For{$i = 1, \ldots, n - 1$} {
                $S^{i + 1} \coloneqq \{\}$\;
                \For{$\rstring \in S^{i}$ and $\text{s} \in \quater$} {
                    $\tilde{\rstring} \coloneqq \text{r}_{1} \cdots \text{r}_{i - 1} \text{s}\, \text{r}_{i + 1} \cdots \text{r}_{n}$\;
                    compute $\cmw_{\tilde{\rstring}}$ according to \eqref{equation:CumulativeWeightsDefinition}\;
                    \If{$\cmw_{\tilde{\rstring}} \neq 0$} {
                        add $\tilde{\rstring}$ to $S^{i + 1}$\;
                    }
                }
            }
            return $\{(\tstring, \cmw_{\tstring} = \omega_{\tstring}) \defcolon \tstring \in S^{n}\}$\;
        \end{algorithm}
    \end{minipage}
\caption{
    Two variants of \texttt{TPD}.
    The recursive version inputs a matrix $A$, a list $L$ for storing the quaternary strings and associated weights, and a current (possibly incompletely expanded) quaternary string $\sstring$.
    If $A$ already is a scalar it entails, by constructing, the weight $\omega_{\sstring}$ for the current string $\sstring$ which is therefore appended to $L$.
    Otherwise the CMWs are computed via matrix slicing, and \texttt{TPD}($\cmw_{\text{t} \bm{\ast}}$, $L$, $\text{t} \sstring$) for each nonzero CMW $\cmw_{\text{t} \bm{\ast}}$.
    In the iterative version an initial set of strings $S^{1}$ exclusively contains the wildcard string $\bm{\ast}$.
    In the following we iterate over all qubit positions $i$ except the last one.
    For each such $i$ we initialize an empty set $S^{i + 1}$, expand the $i$-th position of every string $\rstring$ in the predecessor $S^{i}$ with a concrete $\text{s} \in \quater$, calculate the respective CMW, and finally add the altered string $\tilde{\rstring}$ to $S^{i + 1}$ if the corresponding CMW $\cmw_{\tilde{\rstring}}$ is nonzero.
    The returned final set $S^{n}$ therefore includes all entirely expanded quaternary strings $\tstring$ which contribute nonzero weights $\omega_{\tstring}$.
}
\label{figure:Algorithms}
\end{figure*}

\section{Preliminaries}\label{section:Preliminaries}

\subsection{\label{subsection:ProblemFormulation}Problem formulation}

We briefly collect all the relevant basic properties of the Pauli matrices.
Throughout this paper, we denote with $\Mat(d)$ the set of all complex $d \times d$ matrices.
The Pauli matrices $\sigma^{\{0, 1, 2, 3\}} \coloneqq \{I, X, Y, Z\}$ are a set of hermitian, involutory, and unitary $2 \times 2$ matrices.
Tensorizing Pauli matrices with each other leads to \emph{Pauli strings}.
In order to shorten notation, we set $\quater \coloneqq \{0, 1, 2, 3\}$ and define a Pauli string of length $n$ via a corresponding quaternary string $\tstring \in \quater^{n}$ as
\begin{align}\label{equation:PauliString}
    \sigma^{\tstring} \coloneqq \sigma^{\text{t}_{1}} \otimes \sigma^{\text{t}_{2}} \otimes \cdots \otimes \sigma^{\text{t}_{n}}.
\end{align}
The set of Pauli strings $\{\sigma^{\tstring} \defcolon \tstring \in \quater^{n}\}$ again constitutes hermitian, involutory, and unitary matrices which form an orthonormal basis of $\Mat(2^{n})$ with respect to the \emph{Frobenius inner product} $\langle A, B \rangle \coloneqq \frac{1}{2^{n}} \tr(A^{*} B)$.

We address the objective of computing the Pauli decomposition of a given matrix $A \in \Mat(2^{n})$, that is
\begin{align}\label{equation:PauliDecomposition}
    A = \sum_{\tstring \scriptin \quater^{n}} \omega_{\tstring} \sigma^{\tstring},\quad \omega_{\tstring} \coloneqq \tfrac{1}{2^{n}} \tr(\sigma^{\tstring} A).
\end{align}
In the worst case, all $\abs{\quater^{n}} = 4^{n}$ terms contribute which clearly dictates the worst-case scaling for every Pauli decomposition algorithm.
Moreover, the direct calculation of $\omega_{\tstring}$ involves multiplying together two $2^{n} \times 2^{n}$ matrices.

\subsection{\label{subsection:ExistingAlgorithms}Existing algorithms}

Due to the ubiquity of Pauli decomposition, numerous algorithms addressing this task already exist.
We provide a brief summary of those algorithms, all of which are subjected to numerical testing in \autoref{section:Results}, and highlight their worst-case complexities

The H2ZIXY \cite{Pesce2021hH2ZIXYPauliSpinMatrixDecompositionOfRealSymmetricMatrices} algorithm straightforwardly calculates all $4^{n}$ Pauli strings in advance and then formulates a system of $4^{n}$ linear equations, one for each matrix element, to be solved.\footnote{For real-symmetric input matrices, Pauli strings with an odd number of $Y$-matrices can be omitted.}
It yields a worst-case runtime of $\bigo(64^{n})$ as it constructs a linear system of $4^{n}$ equations which is subsequently solved using \texttt{numpy.linalg.solve} which, in turn, uses LAPACK's \texttt{\_gesv} routine.
As the latter operates cubically in the input dimension, i.e. $(4^{n})^{3} = 64^{n}$, we obtain said complexity.

\begin{figure*}[!th]
    \begin{minipage}{0.47\linewidth}
        \begin{algorithm}[H]
            \caption{Partial \texttt{TPD}($A$, $L$, $T$, $\sstring$=``\,'')}
            \eIf{$\dim(A) = 1 \times 1$}
              {
                append $(\sstring, A =\omega_{\sstring})$ to $L$;
              }
              {
                \For{children t of root($T$)}{
                    compute $\cmw_{ \text{t}\bm{\ast}}$ according to \eqref{equation:CumulativeWeightsDefinition}\;
                    \If{$\cmw_{ \text{t}\bm{\ast}} \neq 0$} {
                        call Partial \texttt{TPD}($\cmw_{ \text{t}\bm{\ast}}$, $L$, subtree(t), $\text{t}\sstring$)\;
                    } 
                }
              }
        \end{algorithm}
    \end{minipage}\hfill
    \begin{minipage}{0.47\linewidth}
        \begin{tikzpicture}
    \foreach \i in {0,2,4} {
        \draw (0,\i) to (4,\i);
        \draw (\i,0) to (\i,4);
    }
    \draw (2,2) to (5.5,4) node[fill=white] {$\cmw_{0\bm{\ast}}$};
    \draw (2,2) to (5.5,3) node[fill=white] {$\cmw_{1\bm{\ast}}$};
    \draw (2,2) to (5.5,1.5);
    \draw (2,2) to (5.5,0) node[fill=white] {$\cmw_{3\bm{\ast}}$};
    \node[fill=white] at (2,2) {$M$};
    \foreach \i in {0,1,2} {
        \draw (4.5,0.5+\i) to (6.5,0.5+\i);
        \draw (4.5+\i,0.5) to (4.5+\i,2.5);
    }
    \draw (5.5,1.5) to (7.5,3) node[fill=white] {$\cmw_{20}$};
    \draw (5.5,1.5) to (7.5,2);
    \draw (5.5,1.5) to (7.5,1) node[fill=white] {$\cmw_{22}$};
    \draw (5.5,1.5) to (7.5,0.5) node[fill=white] {$\cmw_{23}$};
    \draw (7,1.5) rectangle ++(1,1);
    \node[fill=white] at (7.5,2) {$\cmw_{21}$};
    \node[draw=black,fill=white] at (5.5,1.5) {$\cmw_{2\bm{\ast}}$};
\end{tikzpicture}
    \end{minipage}
\caption{Algorithm variation with adaptation for partial Pauli decomposition. The additional quaternary tree structure $T$ dictates which CMWs have to be calculated.
After calculating the relevant CMWs for all children of the current tree's root, the Partial \texttt{TPD} is called again with the subtree emerging at the respective children passed as new tree structure.
The right picture depicts an exemplary subtree containing only one path that corresponds to the $\cmw_{21} = \omega_{21}$ weight.}
\label{figure:PartialDecomposition}
\end{figure*}

In contrast, \texttt{SparsePauliOp.from\_operator}, the previous internal Pauli decomposing method of Qiskit \cite{JavadiAbhari2024QuantumComputingWithQiskit} (versions $\leq$ 0.46.1), directly calculated the $4^{n}$ Frobenius inner products of the input matrix with each Pauli string, involving the multiplication of $2^{n} \times 2^{n}$ matrices.\footnote{With the release of version 1.0., Qiskit uses an implementation of our method for \texttt{SparsePauliOp.from\_operator}.}
The underlying algorithm for matrix multiplication may differ between architectures and custom settings, but, in order to be as general as possible, it is fair to assume the standard cubic implementation.
This yields a total worst-case runtime of $\bigo(32^{n})$.

The Pauli Composer \cite{VidalRomero2023PauliComposerComputeTensorProductsOfPauliMatricesEfficiently} algorithm offers a substantial speed-up generating the Pauli strings.
This method relies on the insight that Pauli strings have exactly one non-zero entry per row and column, making it possible to construct them more efficiently in a well-suited sparse format.
By replacing the standard Kronecker product with the Pauli Composer method and harnessing the enhanced efficiency of sparse matrix multiplication, it becomes possible to markedly expedite the computation of Frobenius inner products, improving the worst-case scaling to $\bigo(8^{n})$ for the resulting \emph{Pauli Decomposer} (PD).

In contrast, PennyLane's \cite{Bergholm2022PennyLaneAutomaticDifferentiationOfHybridQuantumClassicalComputations} internal Pauli decomposition routine follows an entirely distinct, quantum-inspired approach:
The weights $\omega_{\tstring}$ are inferred via Bell measurements on the pure Choi matrix of the superoperator $\rho \mapsto A \rho A^{*}$.
This approach is also elaborated on in more detail by \citet{Hamaguchi2023HandbookForEfficientlyQuantifyingRobustnessOfMagic}, including a derivation of its $\bigo(n 4^{n})$ worst-case scaling.

Finally, a very recent approach by \citet{Jones2024DecomposingDenseMatricesIntoDensePauliTensors} leverages the diagonal/anti-diagonal structure of Pauli matrices as well as Gray codes to find another favorable way of computing the weights $\omega_{\tstring}$, also admitting a worst-case runtime of $\bigo(8^{n})$.

\section{Methods}\label{section:Methods}

Our approach is centered around expanding the decomposition \eqref{equation:PauliDecomposition} in its tensor factors and calculating their weights via matrix partitioning rather than multiplication.
Namely, by equally partitioning a given input matrix $A \in \Mat(2^{n})$ into four blocks of dimension $2^{n -1} \times 2^{n - 1}$, we obtain the essential relation
\begin{align}\label{equation:TensorizedDecomposition}
    A &= \begin{bmatrix}
    A_{11} & A_{12} \\
    A_{21} & A_{22} \\
    \end{bmatrix}= \sum_{\text{t} \scriptin \quater} \sigma^{\text{t}} \otimes \cmw_{\text{t} \bm{\ast}} ,
\end{align}
with the \emph{cumulative matrix weights} (CMW)
\begin{align}\label{equation:CumulativeWeightsDefinition}
\begin{split}
    \cmw_{0 \bm{\ast}} &= \tfrac{1}{2} (A_{11} + A_{22}),\ \cmw_{1 \bm{\ast}} = \tfrac{1}{2} (A_{12} + A_{21}),\\
    \cmw_{2 \bm{\ast}} &= \tfrac{i}{2}(A_{12} - A_{21}),\ \cmw_{3 \bm{\ast}} = \tfrac{1}{2}(A_{11} - A_{22}).
\end{split}
\end{align}

The calculation of $\cmw_{\text{t} \bm{\ast}}$ does not involve any matrix multiplication, is therefore comparatively cheap, and entails, by construction, the weighted sum of all Pauli strings having $\sigma^{\text{t}}$ in their first tensor factor, i.e.,
\begin{align}\label{equation:CumulativeWeightsFormula}
    \cmw_{\text{t}\ bm{\ast}} = \sum_{\sstring \scriptin \quater^{n - 1}} \omega_{\text{t} \sstring} \sigma^{\sstring}
\end{align}
Most importantly, if $\cmw_{\text{t} \bm{\ast}} = 0$, the multi-linearity of the tensor product yields that $\omega_{\tstring'} = 0$ for all $\tstring' \in \quater^{n}$ with $\text{t}'_{1} = \text{t}$.
Hence, in this case we would have already determined the values of $\lvert \{\text{t}\} \times \quater^{n - 1}\rvert = 4^{n - 1}$ weights.
This also means that additional input structures like diagonality or symmetry which rule out a large subset of Pauli strings are detected early.

In an iterative manner, we now equally partition each nonzero CMW into four blocks as well, yielding analogues of \eqref{equation:TensorizedDecomposition} and \eqref{equation:CumulativeWeightsDefinition} with new CMWs
\begin{align}\label{equation:CumulativeWeightsFormula2}
    \cmw_{\text{t} \text{s} \bm{\ast}} = \sum_{\rstring \scriptin \quater^{n - 2}} \omega_{\text{t} \text{s} \rstring} \sigma^{\rstring}
\end{align}
which can be further partitioned to yield new CMWs and so on.
The scheme continues until the remainder strings $\bm{\ast}$ are entirely expanded into concrete strings $\tstring \in \quater^{n}$, respectively.
Correspondingly, in the last iteration step, the CMWs become scalar-valued and yield the remaining individual weights $\omega_{\tstring}$.
An iterative and a recursive version of this \emph{Tensorized Pauli Decomposition} (\texttt{TPD}) are summarized as pseudocode in \autoref{figure:Algorithms}.

In practice, one is often given a guarantee on which Pauli strings could contribute in a decomposition, or, algorithmically equivalently, one may only be interested in certain weights $\omega_{\tstring}$ in the first place.
This additional requirement can be represented as a collection of paths within a complete quaternary tree with $n$ layers;
each path corresponds to exactly one Pauli string of interest.
The recursive variant of \texttt{TPD} is especially suited for this adaptation:
We further input a quaternary subtree $T$ entailing the collection of paths/Pauli strings in question.
For each children of the root node, we calculate the associated CMW, register its corresponding index $t \in \quater$, and call the routine with the obtained CMW and the subtree emerging at the picked child node.
The entire algorithmic procedure is summarized as pseudocode in \autoref{figure:PartialDecomposition}.

\begingroup
  \hypersetup{hidelinks}
\begin{table*}[!ht]
    \begin{minipage}{0.48\linewidth}
            \begin{tabular}{|c|c|c|} \hline
                algorithm & \hspace*{0.25cm}worst-case\hspace*{0.25cm} & \hspace*{0.25cm}TFIM\hspace*{0.25cm} \\\hline
                \texttt{TPD}, \hyperlink{cite.JavadiAbhari2024QuantumComputingWithQiskit}{Qiskit} (version $\geq 1.0$) & $\bigo(n 4^{n})$ & $\bigo(n 2^{n})$ \\
                \hyperlink{cite.Bergholm2022PennyLaneAutomaticDifferentiationOfHybridQuantumClassicalComputations}{PennyLane}, \hyperlink{cite.Hamaguchi2023HandbookForEfficientlyQuantifyingRobustnessOfMagic}{Hamaguchi \textit{et al.}} & $\bigo(n 4^{n})$ & $\bigo(n 4^{n})$ \\
                \hyperlink{cite.VidalRomero2023PauliComposerComputeTensorProductsOfPauliMatricesEfficiently}{Pauli Decomposer}, \hyperlink{cite.Jones2024DecomposingDenseMatricesIntoDensePauliTensors}{Jones} & $\bigo(8^{n})$ & $\bigo(8^{n})$ \\
                \hyperlink{cite.JavadiAbhari2024QuantumComputingWithQiskit}{Qiskit} (version $\leq 0.46.1$) & $\bigo(32^{n})$ & $\bigo(32^{n})$ \\
                \hyperlink{cite.Pesce2021hH2ZIXYPauliSpinMatrixDecompositionOfRealSymmetricMatrices}{H2ZIXY} & $\bigo(64^{n})$ & $\bigo(64^{n})$ \\
                \hline
            \end{tabular}
            \caption{\label{table:AlgorithmComparison}Asymptotic performances for the \texttt{TPD} and other existing decomposition routines in the worst-case and for the $n$-qubit TFIM Hamiltonian.
            The \texttt{TPD} and PennyLane's method share the same worst-case runtime; the other algorithms perform asymptotically worse in descending order.
            The \texttt{TPD} manages to significantly reduce the runtime for the TFIM use-case, provided that it is not known in advance that the input is a TFIM Hamiltonian.
            In contrast, for all other algorithms, the TFIM decomposition -- again without prior knowledge -- is asymptotically as costly as the general worst case.}
    \end{minipage}\hfill
    \begin{minipage}{0.48\linewidth}
            \begin{tabular}{|c|c|}\hline
                cases & \texttt{TPD} scaling \\\hline
                general worst-case & $\bigo(n 4^{n})$ \\
                TFIM Hamiltonian & $\bigo(n 2^{n})$ \\
                single Pauli string & $\bigo(2^{n})$ \\
                $p$ Pauli strings, $m$ matrix entries & $\bigo(m n p)$ \\
                $p$ Pauli strings & $\bigo(2^{n} p^{3 / 2})$ \\
                $m$ matrix entries & $\bigo(m^{2} n 2^{n})$ \\
                \hline
            \end{tabular}
            \caption{\label{table:TPDSpecialCases}Asymptotic \texttt{TPD} performances for several cases: worst-case, TFIM Hamiltonian, single Pauli string, prescribed number of Pauli strings and/or nonzero matrix elements.
            The runtimes for known number of nonzero matrix elements $m$ are especially relevant for applications where this quantity is easy to determine, while the complexities in terms of Pauli strings $p$ are both relevant for initial guarantees on the contributing Pauli strings and for the partial Pauli decomposition.}
    \end{minipage}
\end{table*}
\endgroup

\section{Results}\label{section:Results}

As already mentioned in \autoref{section:Preliminaries}, the worst case constitutes $4^{n}$ contributing terms which all have to be calculated;
this factor can, in full generality, never be eliminated.
However, additional factors vary drastically among different algorithms.
The \texttt{TPD} -- whether iterative or recursive -- matches the best existing worst-case scaling:
In each iteration $i$, $4^{i}$ CMWs are calculated which each involve taking the sum/difference of two $2^{n - i} \times 2^{n - i}$ matrices respectively and the subsequent element-wise division by two, hence $\bigo(4^{n})$ operations per iteration.
Executing all $n$ iterations, this yields a worst-case runtime of $\bigo(n 4^{n})$.

A more detailed analysis yields runtimes for some special cases.
For a given input, let $m$ be the number of non-zero matrix entries and $p$ be the number of contributing Pauli strings.
Then the maximum width of the quaternary subtree explored by the \texttt{TPD} -- and thus the maximum number of CMWs that have to be computed in the same iteration -- is upper bounded by $p$.
Furthermore, the number of nonzero entries in any CMW cannot be higher than the number of nonzero entries in its parent matrix.
Therefore, via an inductive argument, the number of additions performed to calculate any CMW is upper bounded by $m$.
In summary, in each iteration, $m p$ operations have to be performed, yielding, in total, a scaling of $\bigo(m n p)$.

If we only prescribe the number of qubits $n$ and the number of contributing Pauli strings $p$, we can use the fact that every Pauli string constitutes at most $2^{n}$ nonzero matrix elements \cite{VidalRomero2023PauliComposerComputeTensorProductsOfPauliMatricesEfficiently} to get an estimate of $m \sim p 2^{n}$ for the initial number of nonzero matrix elements.
This argument can be repeated for every iteration $i$, where we decompose an $(n - i)$-qubit matrix, yielding at most $p 2^{n - i}$ nonzero matrix elements per CMW in the $i$-th iteration.
For the first $\lceil\log_{4} p\rceil$ iterations $i$, we can additionally upper bound the number of simultaneous calculations of CMWs by $4^{i}$, respectively.
Therefore, their total costs amount to
\begin{align*}
    \sum_{i = 1}^{\lceil\log_{4} p\rceil} 4^{i} 2^{n - i + 1} p = 2^{n + 2} p (p^{1 / 2} - 1) \in \bigo(2^{n} p^{3 / 2}).
\end{align*}
For the remaining $n - \lceil\log_{4} p\rceil$ iterations, the number of simultaneously considered CMWs is trivially upper bounded by $p$.
Therefore, we incur additional costs of
\begin{align*}
    \sum_{i = \lceil\log_{4} p\rceil}^{n} p^{2} 2^{n - i + 1} = 2^{n + 2} p^{2} (p^{-1 / 2} - 1) \in \bigo(2^{n} p^{3 / 2}),
\end{align*}
yielding, in total, a scaling of $\bigo(2^{n} p^{3 / 2})$.

Conversely, the fact that every Pauli string contributes $\bigo(2^{n})$ nonzero matrix elements implies that only $\bigo(2^{n})$ Pauli strings can contribute to the same matrix entry – if more would contribute there would not be enough Pauli strings left for the remaining matrix elements – yielding an estimate $p \sim m 2^{n}$ in case only $m$ and $n$ are prescribed.

\begin{figure*}
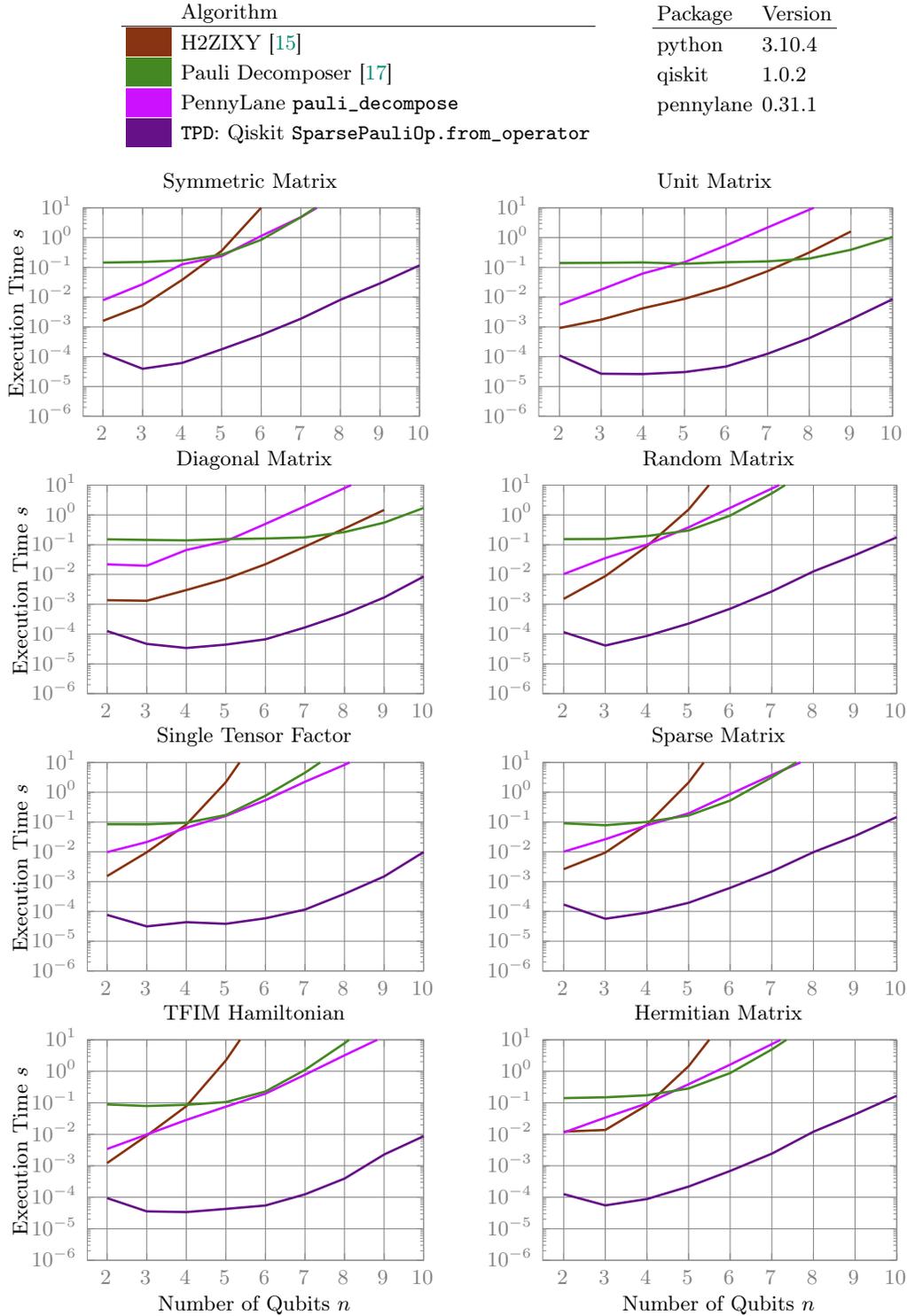

    \definecolor{h2}{HTML}{883311}
    \definecolor{pl}{HTML}{CC11FF}
    \definecolor{Qk}{HTML}{661188}
    \definecolor{PD}{HTML}{448822}

		\hspace*{6pt}\raisebox{0\height}{
		\begin{tabular}{l l}
			  & Algorithm \\\hline
			\colorbox{h2}{\phantom{dot}} & H2ZIXY \cite{Pesce2021hH2ZIXYPauliSpinMatrixDecompositionOfRealSymmetricMatrices} \\
			\colorbox{PD}{\phantom{dot}} & Pauli Decomposer \cite{VidalRomero2023PauliComposerComputeTensorProductsOfPauliMatricesEfficiently} \\
			\colorbox{pl}{\phantom{dot}} & PennyLane \texttt{pauli\_decompose}\\
			\colorbox{Qk}{\phantom{dot}} & \texttt{TPD}: Qiskit \texttt{SparsePauliOp.from\_operator} \\
		\end{tabular}
		} 
        \hspace*{15pt}\raisebox{0.2\height}{
        \begin{tabular}{l l}
            Package & Version\\ \hline
        	python & 3.10.4 \\
        	qiskit & 1.0.2 \\
        	pennylane & 0.31.1
        \end{tabular}
        }
        \\
        \vspace*{2pt}
        \input{symm.tex}
        \input{unit.tex} \\
        \vspace*{-10pt}
        \hspace*{0.45pt}
	\input{diag.tex}
        \input{rand.tex} \\
        \vspace*{-10pt}
        \hspace*{0.45pt}
        \input{oneT.tex}
        \input{spars.tex} \\
        \vspace*{-10pt}
        \hspace*{0.45pt}
        \input{tfim.tex}
        \input{herm.tex}
        
        \vspace*{-1pt}
     \caption{Measured execution times of the various Pauli decompositions of different matrix types.
     In each plot, the execution times are drawn for one matrix type, showing qubit number, and the execution times on a logarithmic axis.
     PennyLane's decomposition method yields the same moderate runtime for all input examples, showcasing its insensibility to any special structure of the input.
     With the exception of the unit matrix and diagonal matrices, the H2ZIXY algorithm performs the poorest;
     in these two specific cases, however, it is able to outperform PennyLane's method.
     The Pauli Decomposer is, in each case, compatible with the best performing method besides \texttt{TPD}, especially yielding favorable runtimes for when the input is highly structured.
     Lastly, Qiskit's implementation of the \texttt{TPD} outperforms all other algorithms on every single benchmark set.
     It also shows favorable runtimes for highly structured input.}
    \label{figure:Numerics}
\end{figure*}

To further validate the effectiveness of our algorithm, we assess our algorithm's performance in decomposing Hamiltonians into the Pauli basis.
Specifically, we consider the \texttt{TPD}'s performance on the symmetric one-dimensional Transverse Field Ising Model (TFIM) Hamiltonian \cite{Stinchcombe1973IsingModelInATransverseFieldBasicTheory}
\begin{align}\label{equation:TFIMHamiltonian}
    H_{n} = - J \bigg(\sum_{i = 1}^{n - 1} \sigma_{i}^{3} \sigma_{i + 1}^{3} + g \sum_{j = 1}^{n} \sigma_{j}^{1}\bigg),
\end{align}
a significant benchmark commonly used in quantum computing.\footnote{The TFIM is a quantum spin model on a one-dimensional lattice where spins interact with neighboring spins and are subject to an external magnetic field perpendicular to their orientation.}
Obviously, \eqref{equation:TFIMHamiltonian} already entails the entire decomposition into $2 n - 1$ Pauli strings, so one has to be careful on how to meaningfully compare the performance of different algorithms on this example.
Our understanding is the following:
Conducting a specialized runtime analysis is only possible if all the relevant structure of the input is known.
In the case of Pauli decomposition, this relevant structure effectively matches the output of the algorithm itself.
Therefore, a detailed performance analysis is only possible for ``solved'' instances such as the TFIM Hamiltonian.
However, one has to discriminate between information we assume for the runtime analysis and information we assume for executing the algorithm.
In fact, if we assume to know in advance that the input is a TFIM Hamiltonian and then execute e.g. a version of the Pauli Decomposer tailored to the partial decomposition with exactly the Pauli strings that appear within the TFIM Hamiltonian, we obtain a runtime of $\bigo(n 2^{n})$;
but this is not what we are comparing against.
In contrast, we assume no prior knowledge when executing the algorithm.
Accordingly, a sophisticated partial decomposition scheme cannot be given information on which Pauli weights to compute, and e.g. the Pauli Decomposer falls back to its worst-case complexity of $\bigo(8^{n})$.
In the following, we show that the \texttt{TPD} still obtains an improved runtime that automatically arises from the (a priori unknown) input structure.

As a prerequisite, we consider the case of a single Pauli string first.
Starting from $m = 2^{n}$ nonzero matrix elements, each iteration halves the number of nonzero matrix elements in the current CMW, meaning that in iteration $i$ there are $2^{n - i}$ additions to be performed.
Adding up all operations over the course of $n$ iterations thus yields a scaling of $\bigo(2^{n})$.
Moving to $H_{n}$, the following decomposition yields valuable insights into the runtime:
\begin{align}\label{equation:FirstIterationTFIMHamiltonian}
    H_{n} = \sigma^{0} \otimes H_{n - 1} - J \sigma_{1}^{3} - J g \sigma_{1}^{1}.
\end{align}
Since $H_{n}$ has linearly many contributing Pauli strings, there are $\bigo(n 2^{n})$ nonzero matrix elements, dictating the cost of the first iteration.
Now, \eqref{equation:FirstIterationTFIMHamiltonian} shows that the remaining iterations accumulate to the cost of decomposing $H_{n - 1}$ as well as two single Pauli strings, the later is $\bigo(2^{n})$.
Let $\bigo(H_{n})$ denote the decomposition scaling for $H_{n}$, then we obtain the recursion relation
\begin{align}\label{equation:TFIMRecursiveRelation}
    \bigo(H_{n}) = \bigo(H_{n - 1}) + \bigo(n 2^{n}).
\end{align}
Successively expanding the $\bigo(H_{i})$ terms for all $i < n$ eventually yields an overall scaling of $\bigo(n 2^{n})$.

We emphasize that the speed-up $\bigo(n 4^{n}) \to \bigo(n 2^{n})$ for the TFIM Hamiltonian is a unique property of the \texttt{TPD} in comparison to the other methods.
The only promise that can be made is that $H_{n}$ is real-symmetric;
of the existing algorithms mentioned, only the H2ZIXY algorithm and the Pauli Composer are sensible for that kind of additional input information.
They are then able to omit taking into account all Pauli string with an odd number of $\sigma^{2}$-operators.
However, as this reduced the number of relevant Pauli strings from $4^{n}$ to $2^{n - 1} (2^{n + 1})$, the asymptotic scaling stays the same.
An overview over the general worst-case complexity as well as for the special case of the TFIM Hamiltonian is provided in \autoref{table:AlgorithmComparison}, showcasing the \texttt{TPD}'s favorable scaling in comparison to the other methods.
Furthermore, the calculated \texttt{TPD} complexities for all other special cases are collected in \autoref{table:TPDSpecialCases}.

We now address the \texttt{TPD}'s space complexity.
In each iteration $i$, up to $4^{i}$ matrices of dimension $2^{n - i} \times 2^{n - i}$ are created which would introduce $\bigo(4^{n})$ additional memory per iteration, i.e. $\bigo(n 4^{n})$ additional memory in total.
However, since information is only passed between directly consecutive iterations, the parent matrices' memory can be reused after finishing each iteration, yielding a milder memory overhead of $\bigo(4^{n})$.
This is in alignment with the space complexity of PennyLane's method \citep{Hamaguchi2023HandbookForEfficientlyQuantifyingRobustnessOfMagic}.
The Pauli Decomposer works with a spatial overhead of $\bigo(2^{n})$ which is improved by \citep{Jones2024DecomposingDenseMatricesIntoDensePauliTensors} to a constant overhead.

Algorithms with asymptotically better worst-case or even special case complexities do not necessarily perform well in practice;
numerical studies remain essential.
In order to also cover this issue, we further investigate numerically the performance of \texttt{TPD} against the H2ZIXY, the Pauli Decomposer method, and PennyLane's internal implementation on instances from 2 to 10 qubits.
The results are depicted in \autoref{figure:Numerics} and include input matrices of specific types: symmetric, diagonal, random, sparse, single non-trivial tensor factor, and hermitian matrices as well as the unit matrix and the TFIM Hamiltonian, respectively.
All matrices except the TFIM Hamiltonian as well as the unit matrix are created using matrix elements drawn uniformly at random from the interval $[0, 1]$.
Symmetric and Hermitian matrices are manipulated accordingly to guarantee their properties.

In all numerical experiments except for diagonal and unit matrices, the H2ZIXY algorithm exhibits the poorest performance.
The implementation in PennyLane consistently exhibits identical behavior across all scenarios, showcasing the methods' insensitivity to predefined characteristics of the input data.
The Pauli Decomposer exhibits runtime performance comparable to the aforementioned algorithms but significantly surpasses them in the case of diagonal and unit matrices.

Lastly, Qiskit's implementation of the \texttt{TPD} consistently outperform all its competitors, with a substantial performance advantage becoming evident as the number of qubits increases.

\section{Conclusion and Outlook}\label{section:ConclusionAndOutlook}

The task of computing the Pauli decomposition of multi-qubit operators is at the core of many applications from quantum computing, including Hamiltonian simulation and quantum error correction.
In this article we have proposed a simple, yet powerful method for accelerating this ubiquitous task.
Our method, the Tensorized Pauli Decomposition (\texttt{TPD}) algorithm avoids expensive matrix multiplication and, instead, calculates the Pauli weights in an recursive/iterative manner via matrix slicing.

Our detailed complexity analysis revealed an optimal worst-case scaling of the \texttt{TPD} among existing methods.
For several special cases, we concluded the \texttt{TPD}'s drastically improved scaling, unmatched by the other methods.
Furthermore, we have tested the \texttt{TPD} against the most popular alternatives on various instances from $2-10$ qubits and observed a decreased runtime in favor of the \texttt{TPD} in all cases.
Based on the observed trend in our numerical experiments as well as the asymptotically favorable scaling, we anticipate improved algorithmic efficiency for most instances across the respective domains.
By detecting the input's structure early on, \texttt{TPD} achieves speed-ups even in cases where its competitors receive additional flags, lowering their worst-case runtime.

The implementation of the \texttt{TPD} within Qiskit already covers various optimization techniques that are not addressed in this paper, significantly contributing to the achieved speed-up observed in the numerical experiments.
Further investigations of structurally similar tasks may extend the applicability of our algorithmic primitives and of these optimization techniques.
Natural candidates for similar decomposition techniques are other prominent matrix bases which consist of tensorized components such as (generalized) Gell-Mann matrices or Weyl operators.

\begin{acknowledgments}

We thank Tobias J.\ Osborne and Luis Santos for helpful discussions, and Tyson Jones for additional numerical studies and more in-depth discussions. 
Furthermore, we thank Jake Lishman for testing and improving the implementation and integrating it into Qiskit.
LB acknowledges financial support by the Quantum Valley Lower Saxony.

\noindent\textbf{Data and code availability statement.}
The depicted data is available upon reasonable request from the authors.
The Qiskit-independent source code is available on GitHub: \url{https://github.com/HANTLUK/PauliDecomposition}.

\end{acknowledgments}

\twocolumngrid

\bibliographystyle{apsrev4-2}
\bibliography{main.bib}

\end{document}